\documentclass[11pt]{article}
\newcommand{\Comment}[1]{{}}
\usepackage{amsfonts,amsthm,amsmath,amssymb,slashed}
\usepackage[textwidth = 430 pt, textheight = 630 pt]{geometry}
\usepackage{color}

\Comment{\usepackage{color}
\definecolor{MyDarkBlue}{rgb}{0.15,0.15,0.45}
\usepackage[linktocpage=true]{hyperref}
\hypersetup{
colorlinks=true,
citecolor=MyDarkBlue,
linkcolor=MyDarkBlue,
urlcolor=MyDarkBlue,
pdfauthor={Horatiu Nastase and Ulisses M. Portugal},
pdftitle={The monopole problem in holographic cosmology},
pdfsubject={hep-th}
}

\usepackage[numbers,sort&compress]{natbib}
\usepackage{hypernat}}
\usepackage{graphicx}
\usepackage{cite}

\newcommand\ignore[1]{}
\def\one{{\,\hbox{1\kern-.8mm l}}}

\def\Tr{{\rm Tr\, }}

\def\Tr{\mathop{\rm Tr}\nolimits}

\newcommand{\Cset}{{\,\,{{{^{_{\pmb{\mid}}}}\kern-.45em{\mathrm C}}}}}

\newcommand{\be}{\begin{equation}}
\newcommand{\bea}{\begin{eqnarray}}

\newcommand{\ee}{\end{equation}}
\newcommand{\eea}{\end{eqnarray}}

\begin{document}

\renewcommand{\thefootnote}{\fnsymbol{footnote}}

\makeatletter
\@addtoreset{equation}{section}
\makeatother
\renewcommand{\theequation}{\thesection.\arabic{equation}}

\rightline{}
\rightline{}
%   \vspace{1.8truecm}

%\begin{flushright}
% preprint nrs.
%\end{flushright}

%\vspace{10pt}

%%%%%%%%%%%%%%%%%

\begin{center}
{\LARGE \bf{\sc The monopole problem in holographic cosmology}}
\end{center} 
 \vspace{1truecm}
\thispagestyle{empty} \centerline{
{\large \bf {\sc Horatiu Nastase${}^{a}$}}\footnote{E-mail address: \Comment{\href{mailto:horatiu.nastase@unesp.br}}{\tt horatiu.nastase@unesp.br}}
{\bf{\sc and}}
{\large \bf {\sc Ulisses M. Portugal${}^{a}$}}\footnote{E-mail address: \Comment{\href{mailto:ulisses.portugal@unesp.br}}{\tt 
ulisses.portugal@unesp.br}}
                                                        }

\vspace{.5cm}

%\vspace{.3cm}

\centerline{{\it ${}^a$Instituto de F\'{i}sica Te\'{o}rica, UNESP-Universidade Estadual Paulista}} 
\centerline{{\it R. Dr. Bento T. Ferraz 271, Bl. II, Sao Paulo 01140-070, SP, Brazil}}
%\vspace{.3cm}
%\centerline{{\it ${}^b$STAG Research Center and Mathematical Sciences, University of Southampton,}}
%\centerline{{\it Highfield, Southampton SO17 1BJ, United Kingdom}}

\vspace{1truecm}

%%%%%%%%%%%%%%%%%
\thispagestyle{empty}

\centerline{\sc Abstract}

\vspace{.4truecm}

\begin{center}
\begin{minipage}[c]{380pt}
{\noindent In this letter we clarify that the monopole problem can always be solved in bosonic holographic 
cosmology, by the analogue of "dilution" in inflation, which is the fact that the {\em electric} current is an 
irrelevant operator in the dual field theory. We show that not only specific toy models
solve the problem, but any purely bosonic member of the phenomenological class, of 
super-renormalizable, generalized conformal symmetric models.
}
\end{minipage}
\end{center}

\vspace{.5cm}

\setcounter{page}{0}
\setcounter{tocdepth}{2}

\newpage

%\tableofcontents
\renewcommand{\thefootnote}{\arabic{footnote}}
\setcounter{footnote}{0}

\linespread{1.1}
\parskip 4pt

%{}~
%{}~

%---------------------------------------------------------

%%%%%%%%%%%%%%%%%%%%%%%%%%%%%%%%%%%%%%%%%%%%%%%%%%%%%%%%%%%%%%%%%%%%%%%%%%%%%%%%%%%%%%%%
\section{Introduction}
%%%%%%%%%%%%%%%%%%%%%%%%%%%%%%%%%%%%%%%%%%%%%%%%%%%%%%%%%%%%%%%%%%%%%%%%%%%%%%%%%%%%%%%%

The standard paradigm of cosmology, $\Lambda$CDM plus inflationary cosmology
\cite{Brout:1977ix, Starobinsky:1979ty, Starobinsky:1980te, Sato:1980yn, Guth:1980zm, Linde:1981mu, Albrecht:1982wi}, 
matches all known experimental data (after suitable fitting of 
the parameters coming from specific models) and solves all problems: specifically the CMBR fluctuation data can be fitted, the standard
pre-inflationary Hot Big Bang cosmology problems can be solved \cite{Guth:1980zm, Linde:1981mu, Albrecht:1982wi}, 
and one can generate the Standard Model particles by a model of reheating. 

However, the models of holographic cosmology, based on the application of the AdS/CFT correspondence \cite{Maldacena:1997re}
(see the books  \cite{Nastase:2015wjb,Ammon:2015wua} for more information)
to 3+1 dimensional cosmological backgrounds, specifically the general phenomenological model of McFadden and Skenderis, defined 
in \cite{McFadden:2009fg,McFadden:2010na}, were shown to also fit the same CMBR data and solve the same problems.  
Specifically, in \cite{Afshordi:2016dvb,Afshordi:2017ihr} it was shown that one can make a fit of the phenomenological model to the 
CMBR data with the same number of parameters, that is within 0.5 in $\chi^2$ from the standard $\Lambda$CDM plus inflation fit
(840.0 for holographic cosmology vs. 823.5 for $\Lambda$CDM, see table V in \cite{Afshordi:2017ihr}), 
thus being experimentally indistinguishable from the latter. Further, in \cite{Nastase:2019rsn,Nastase:2020uon} it was shown that the 
usual problems of Hot Big Bang cosmology also also solved by the holographic cosmology models, and in \cite{Nastase:2020fgz}
a model of reheating was proposed. 

However, while the smoothness and horizon problem, entropy problem, perturbations problem and baryon asymmetry 
problem were solved generically, and for the flatness problem the condition on a combination of the parameters, 
$f_1<0$, necessary to have a relevant energy-momentum 
tensor $T_{ij}$ that solves the problem, has a concrete solution, solved on most of the parameter space, for the monopole problem 
only a toy field theory model was analyzed, and shown to solve the problem. 
In this letter we show that we can extend the proof from the toy model to the whole set of purely 
bosonic phenomenological models, first reviewing the solution of the monopole problem by having an electric global symmetry 
current that is an irrelevant operator, then showing how this can be in the generic model. 

\section{Solution of the monopole problem}

First, we quickly review the solution of the monopole problem, based on the fact that the electric current $j_\mu^a$ is an irrelevant operator
(which was found to be true in the toy model analyzed in  \cite{Nastase:2019rsn,Nastase:2020uon}).

The monopole problem is the following. Considering an expanding Universe, with an expanding horizon size, 
undergoing a phase transition, like a Grand Unified Theory (GUT) phase transition. Then the Kibble mechanism guarantees that 
one creates about one monopole per nucleon. The Kibble mechanism refers to the fact that, when the temperature drops and 
the scalar takes a VEV in the vacuum, thus breaking the symmetry, it takes a random value for each causally connected patch. 
Then when taking a 
few patches, one generically has a larger patch with monopole topology for the scalar (thus also for the gauge fields), thus 
creating a monopole per region of the order of the horizon.
When nucleons are formed, a similar mechanism creates one nucleon per horizon, therefore leading to one monopole per nucleon. 

Yet experimentally, direct searches for monopoles on Earth finds less than $10^{-30}$ monopoles per nucleon
\cite{Zeldovich:1978wj} (see also \cite{Weinberg:2008zzc}, chapter 4.1.C), so one needs a mechanism to dilute them. 

In the holographic cosmology model, gauge field perturbations $A_\mu^a$ in the bulk cosmology correspond to global 
non-Abelian symmetry currents $j_\mu^a$ in field theory, and ('t Hooft-Polyakov) monopole configurations in the bulk correspond to 
vortex configurations on the boundary. Dilution of monopoles as cosmological time goes forward corresponds to dilution of the 
vortex current along the inverse RG flow, i.e., towards the UV. Considering the two-point function $\langle \tilde j_\mu^a (p)
\tilde j_\nu^b(-p)\rangle\propto p^{1+2\delta}$, we find that we need the vortex current $\tilde j_\mu^a(p)$ to be a 
(marginally) relevant operator, $\delta(\tilde j)>0$. 

But for conformal theories \cite{Witten:2003ya,Herzog:2007ij} and, by extension, to generalized conformal theories, 
the two-point function of the electric currents $j_\mu^a$, constrained to be of the form
\be
\langle j_\mu^a (p)j_\nu^b(-p)\rangle=\sqrt{p^2}\left(\delta_{\mu\nu}-\frac{p_\mu p_\nu}{p^2}\right)K^{ab}\;,
\ee
implies that the S-dual vortex currents have an inverse $K^{ab}$,
\be
\langle \tilde j_\mu^a (p)\tilde j_\nu^b(-p)\rangle=\sqrt{p^2}\left(\delta_{\mu\nu}-\frac{p_\mu p_\nu}{p^2}\right)K^{-1}_{ab}\;,
\ee
so in effect an inverse anomalous dimension, $\delta(j)=-\delta(\tilde j)$. In other words, the need for a (marginally) relevant 
vortex current $\tilde j_\mu^a$ implies the need for a (marginally) irrelevant electric current $j_\mu^a$. 

Finally,  in \cite{Nastase:2019rsn,Nastase:2020uon} it was noted that one has to check that a vortex solution does exist, so that 
the vortex current exists also. Yet, since the 't Hooft-Polyakov monopole in the bulk would be dual to a non-Abelian vortex in 
field theory, and such non-Abelian vortices were found only in complicated theories, not relevant for the phenomenological action, 
with all the fields in the adjoint of $SU(N)$, Dirac monopoles (singular, point-like) were considered instead in the bulk, corresponding 
in the field theory on the boundary to what was dubbed "Dirac vortices", namely singular, point-like versions of the vortices. 
The toy model considered in \cite{Nastase:2019rsn,Nastase:2020uon} admitted such solutions, so there exist models within 
the phenomenological class with vortex solutions. Since we will deal with the whole phenomenological class,
and the existence of cosmological monopoles, thus of field theory vortices, must be imposed on any consistent model, it is enough to know 
that such models exist.

\section{The electric current $j_\mu^a$ is an irrelevant operator}

In this section, we will review the general points of the calculation of the electric current two-point 
function in \cite{Nastase:2019rsn,Nastase:2020uon}), and point out that they apply to any model in the phenomenological class
that has a global symmetry current. 

The phenomenological class, written in the form with 4-dimensional dimensions for the fields, has an Euclidean action
for one gauge field $A_\mu$ and scalar $\Phi^M$ and fermion $\psi^L$ fields in the adjoint of $SU(N)$, given by
\bea
S_{\rm QFT}&=&\frac{1}{g^2_{YM}}\int d^3x \Tr\left[\frac{1}{2}F_{ij}F^{ij}+\delta_{M_1M_2}D_i\Phi^{M_1}D^i\Phi^{M_2}
+2\delta_{L_1L_2}\bar \psi^{L_1}\gamma^i D_i
\psi^{L_2}\right.\cr
&&\left.+\sqrt{2}\mu_{ML_1L_2}\Phi^M \bar \psi^{L_1}\psi^{L_2}+\frac{1}{6}
\lambda_{M_1...M_4}\Phi^{M_1}...\Phi^{M_4}\right]\;,\label{phenoaction}
\eea
plus a nonminimal coupling of gravity to the scalar $1/(2g^2_{YM})\int \xi_M R(\Phi^M)^2$. It has dimensionless couplings, 
except $g^2_{YM}$, which appears in the quantum theory only in the effective coupling combination 
$g^2_{\rm eff}=\frac{g^2N}{q}$, and the fields have $[A_i]=1=[\Phi^M]$ and $[\psi^L]=3/2$. 
Moreover, we will only consider the case without fermions, so $\psi^L=0$, and the dimensionless couplings 
are $\lambda_{M_1...M_4}, \xi_M$. However, $\xi_M$ only is relevant to the energy-momentum tensor (since otherwise 
the corresponding term is 
zero in a flat background), so the only relevant dimensionless couplings for our calculation are $\lambda_{M_1...M_4}$. 
The model is super-renormalizable, and has generalized conformal symmetry, meaning that the only dimensionful coupling is 
$g_{YM}$. 

Since we have not selected a model with a certain global symmetry within the phenomenological class, we cannot 
write an explicit expression for the global current $j_\mu^a$, as in  \cite{Nastase:2019rsn,Nastase:2020uon}), but we assume 
that one exists. 

\begin{figure}[h]
\begin{center}
\includegraphics[width=140mm]{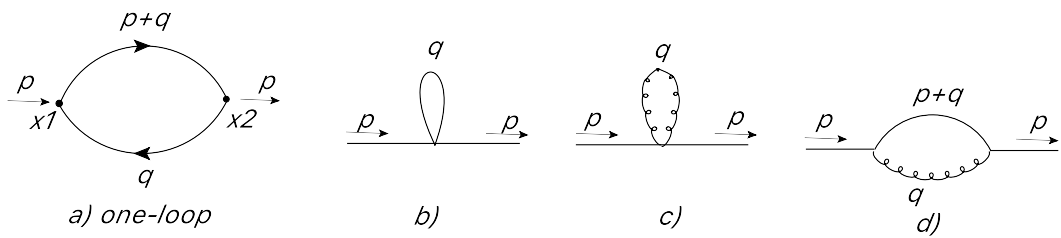}
\end{center}
\caption{One-loop diagrams: a) The unique one-loop diagram for the two-point function of currents. b),c),d) one-loop counterterm
diagrams. }
\label{fig:one-loop}
\end{figure}

\begin{figure}[h]
\begin{center}
\includegraphics[width=80mm]{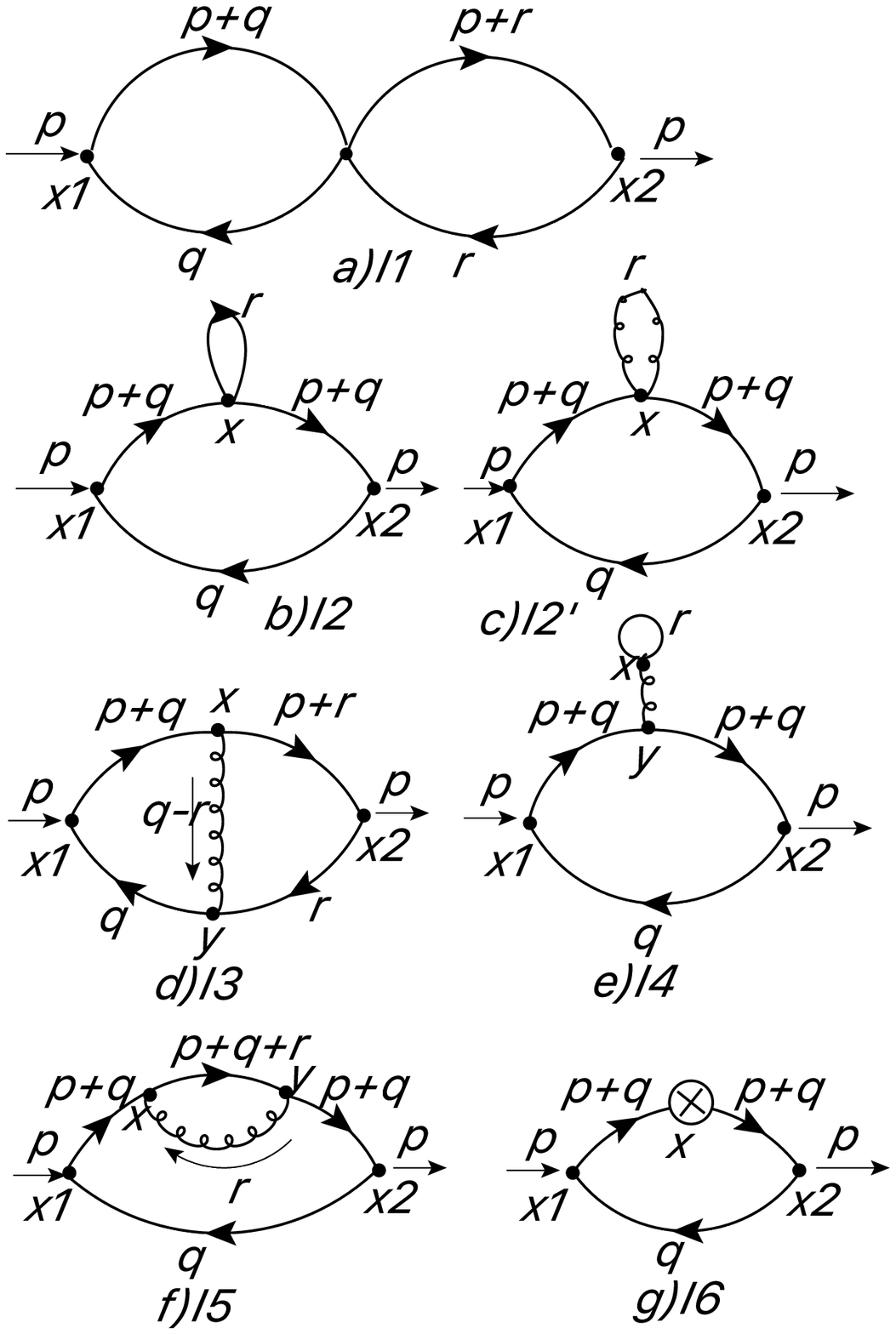}
\includegraphics[width=50mm]{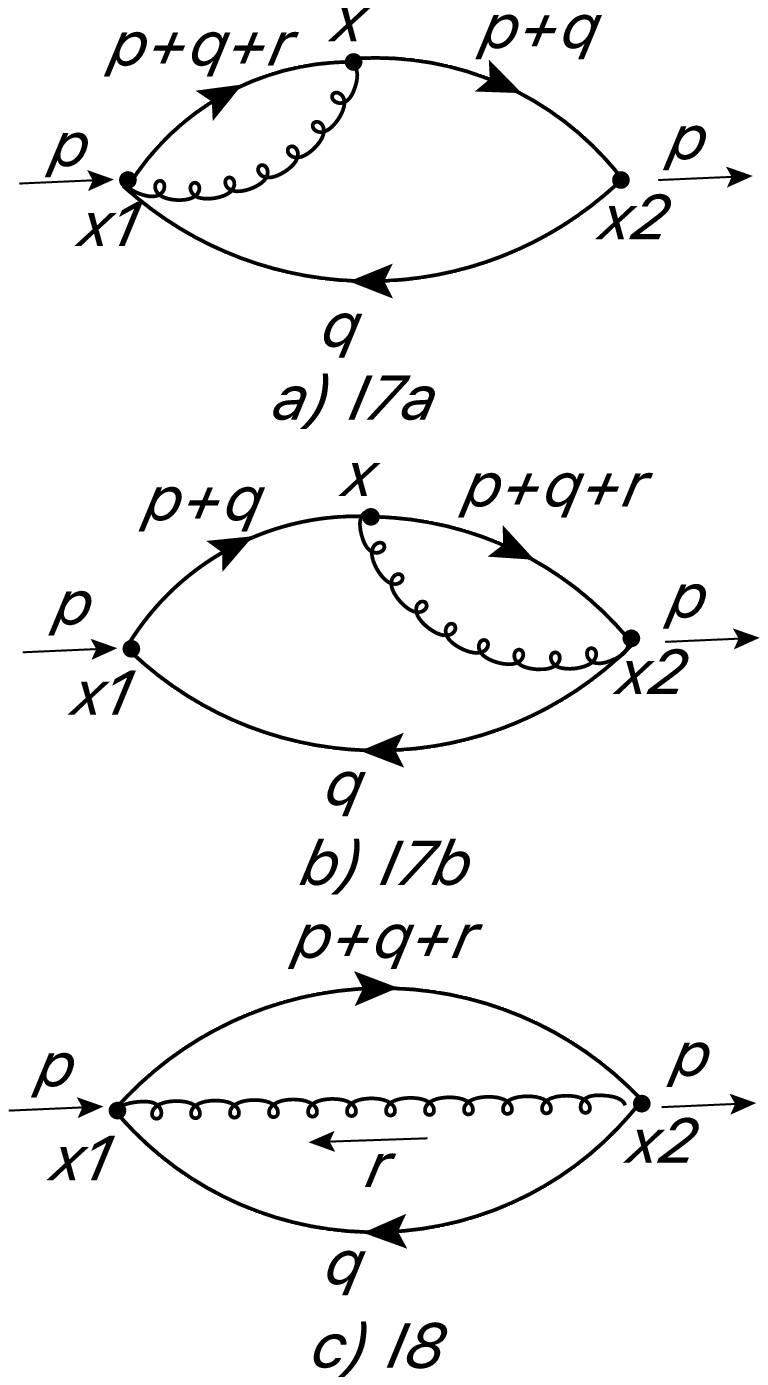}
\end{center}
\caption{Two-loop diagrams without external gauge field insertions (left) and with 
external gauge field insertions (right) ($I_{7a}$ and $I_{7b}$ with one external insertion and $I_8$ with two external 
insertions). }
\label{fig:two-loop}
\end{figure}

To prove that the electric current $j_\mu^a$ is a (marginally) irrelevant operator, we need to calculate 
its two-point function $\langle j_\mu^a (p)j_\nu^b(-p)\rangle$, specifically the finite one-loop result (giving the normalization) 
and the divergent part of the two-loop diagrams (giving the anomalous dimension). 
At one-loop, there is one Feynman diagram for the two-point 
function, and Feynman diagrams for the renormalization, both given in Fig.\ref{fig:one-loop}. At two-loops, there are 
Feynman diagrams with and without external gauge field insertions, in Fig.\ref{fig:two-loop}.

Ignoring the $SU(N)$ indices (which are all taken care of by 't Hooft's double-line notation, as replacing $g^2_{YM}$ with the 
effective $g^2_{\rm eff}=g^2N/q$), the fields of the toy model in \cite{Nastase:2019rsn,Nastase:2020uon} were $\phi_i^a$ and $A_\mu$, 
with 4-point scalar vertices 
\be
-\lambda (2\pi)^3\delta^3(k_1+k_2+k_3+k_4)\epsilon^{abe}\epsilon_{cde}=-2\lambda\delta_{ab}^{cd}(2\pi)^3\delta^3(k_1+k_2+k_3+k_4).
\ee
and 3-point $\phi \phi A$ and 4-point $\phi\phi AA$, as well as pure gluonic ones, completely determined by minimal coupling 
(gauge invariance).  Moreover, by explicit calculation of the Feynman diagrams, one finds that $I_1=I_{2a}=I_{2b}=I_4=0$, 
while the counterterm diagram $I_6$ is finite and multiplied by $\lambda_{\rm ct}$, 
and since the one-loop result is finite, and moreover the theory itself is one-loop finite, so $I_6$ is irrelevant to the calculation as well.

When looking at the calculation of these vanishing integrals, it is easy to see that $I_1,I_{2a}, I_{2b}, I_4$ vanishing is 
unrelated to the form of the vertices: in the case of $I_1$, the 4-point scalar vertex contribution is an overall factor, while for $I_{2a}, I_{2b}, 
I_4$ the vanishing comes from the factorized loop. Also, the fact that $I_6$ is finite is again a result of the fact that it has the topology 
of the one-loop result. So for a generic theory, we are still left with only the diagrams $I_3, I_5, I_{7a}, I_{7b}, I_8$. 
But in these diagrams, since the propagators are still Klein-Gordon (for the scalars and for the gauge field, 
in the Feynman gauge) times delta functions for 
indices, and the relevant vertices are $\phi\phi A$  vertices completely determined by minimal coupling, it is easy to check that the 
expressions for the Feynman diagrams are the same ones written and calculated in \cite{Nastase:2019rsn,Nastase:2020uon}.
The only thing different is an overall factor equal to the number of scalars running in the original scalar loop, but since this factor 
equals the same extra factor at one-loop, this becomes an overall factor in the two-point function, and does not change the value of the 
two-loop anomalous dimension. That means that we still have, for a generic theory, 
\be
\delta(j)=\frac{2}{\pi^2}g^2_{\rm eff}>0\;,
\ee
that is, an irrelevant operator, as we wanted. Therefore, the dual (vortex) current has $\delta (\tilde j)=-\delta(j)<0$, so the 
vortices and holographic dual monopoles get diluted, as we wanted.

\section{Conclusions}

In this letter we have removed a caveat from the previous analysis of the problems with Hot Big Bang cosmology within holographic 
cosmology, showing that the monopole problem is not solved only for some toy models, but generically, for any purely bosonic 
model within the phenomenological class. In terms of matching experimental data and solving theoretical problems, this means that 
the holographic cosmology paradigm is as good as the usual, $\Lambda$CDM plus inflation, paradigm. This should not be a surprise, 
since the two are related by changing the strength of the coupling (inflation means a perturbative gravitational theory, holographic 
cosmology a perturbative 3-dimensional field theory).

{\bf Acknowledgements}

The work of UP is supported in part by FAPESP grant 2018/22008-7.
The work of HN is supported in part by  CNPq grant 301491/2019-4 and FAPESP grants 2019/21281-4 
and 2019/13231-7. HN would also like to thank the ICTP-SAIFR for their support through FAPESP grant 2016/01343-7.

%%%%%%%%%%%%%%%%%%%%%%%%%%%%%%%%%%%%%%%%%%%%%%%%%%%%%%%%%%%%%%%%%%%%%%%%%%%%%%%%%%%%%%%%
\bibliography{HoloCosmoPaper3}
\bibliographystyle{utphys}
%%%%%%%%%%%%%%%%%%%%%%%%%%%%%%%%%%%%%%%%%%%%%%%%%%%%%%%%%%%%%%%%%%%%%%%%%%%%%%%%%%%%%%%%

\end{document}